\documentclass[a4paper,fleqn,usenatbib]{mnras}

\usepackage[T1]{fontenc}
\usepackage{ae,aecompl}

\usepackage{graphicx}	
\usepackage{amsmath}	
\usepackage{amssymb}	

\title[The planetary nebulae K\,4-37]{History of the mass ejection in K\,4-37: from the AGB to the 
evolved planetary nebula phase\thanks{Based on
    observations collected at the German-Spanish 
    Astronomical Center, Calar Alto, jointly operated by the
    Max-Planck-Institut f\"ur Astronomie (Heidelberg) and the Instituto de
    Astrof\'{\i}sica de Andaluc\'{\i}a (CSIC), and on observations acquired
    at the Observatorio Astron\'omico Nacional in the Sierra San Pedro
    M\'artir (OAN-SPM), Baja California, Mexico.}}
\author[L.F. Miranda et al. ]{L.\,F. Miranda$^{1}$\thanks{e-mail:
lfm@iaa.es}, P.\,F. Guill\'en$^{2}$, L. Olgu\'{\i}n$^{3}$, R. V\'azquez$^{2}$ \\
$^{1}$Instituto de Astrof\'{\i}sica de Andaluc\'{\i}a -- CSIC, C/ Glorieta de
la Astronom\'{\i}a s/n, E-18008 Granada, Spain \\
$^{2}$Instituto de Astronom\'{\i}a, Universidad Nacional Aut\'onoma de M\'exico,
Apdo. Postal 877, 22800 Ensenada, B.C., Mexico \\
$^{3}$Departamento de Investigaci\'on en F\'{\i}sica, Universidad de Sonora,
Blvd. Rosales Esq. L.D. Colosio, Edif. 3H, 83190 Hermosillo, Son. Mexico \\
}

\date{Accepted XXX. Received YYY; in original form ZZZ}

\pubyear{2016}

\begin{document}
\label{firstpage}
\pagerange{\pageref{firstpage}--\pageref{lastpage}}
\maketitle

\begin{abstract}
We present narrow-, broad-band, and WISE archive images, and high- and intermediate-resolution 
long-slit spectra of K\,4-37, a planetary nebula that has never been analyzed in detail. 
Although K\,4-37 appears bipolar, the morphokinematical 
analysis discloses the existence of three distinct axes and additional particular directions in the 
object, indicating that K\,4-37 is a multi-axis planetary nebula that has
probably been shaped by several bipolar outflows at different
directions. A 4-6\,M$_{\odot}$ main-sequence progenitor is estimated from the derived high nebular 
He and N abundances, and very high N/O abundance ratio ($\sim$2.32). The general properties are 
compatible with K\,4-37 being a highly evolved planetary nebula located at
$\sim$14\,kpc. The WISE image at 22\,$\mu$m reveals K\,4-37 to be surrounded
by a large ($\sim$13$\times$8\,pc$^2$) elliptical detached shell probably related to 
material ejected from the AGB progenitor. The observed elliptical morphology 
suggests deformation of an originally spherical AGB shell by the ISM magnetic field or 
by the influence of a companion. We compare K\,4-37 and NGC\,6309 
and found remarkable similarities in their physical structure but noticeably different chemical abundances that 
indicate very different progenitor mass. This strongly suggests that, irrespective of the initial mass, their 
(presumably binary) central stars have shared a very similar mass ejection history. 
\end{abstract}

\begin{keywords}
planetary nebulae: individual (K\,4-37) -- insterstellar medium:
jets and outflows.
\end{keywords}



\section{Introduction}

\begin{figure*}
\begin{center}
\includegraphics[width=160mm]{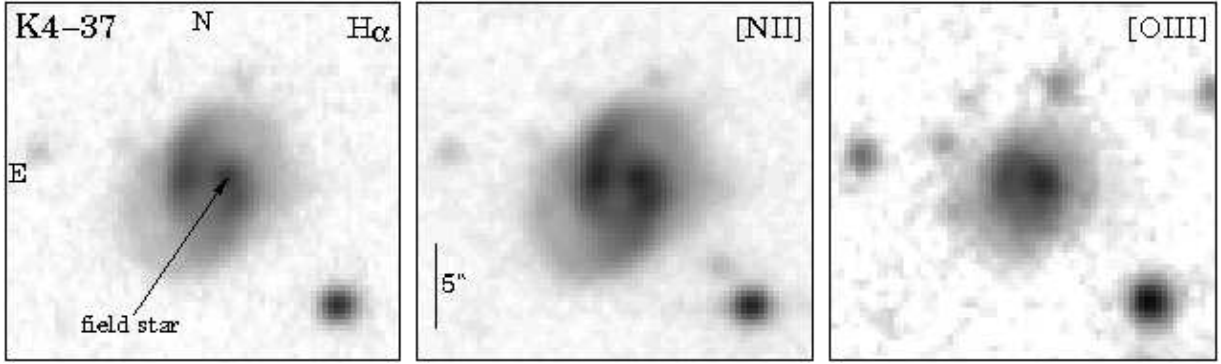}
\caption{Grey-scale reproductions of the 1998 H$\alpha$ and [N\,{\sc ii}], and 2016 [O\,{\sc iii}] images 
of K\,4-37. The grey levels are logarithmic. North is up, east to the left. A field star is arrowed in the 1998 
H$\alpha$ image, which has been identified as a star (and not as a nebular knot)
thanks to the intermediate-resolution, long-slit spectrum (Section\,2.3). The size of the
field shown in the three panels is 23$\times$21.5\,arcsec$^2$.}
\end{center}
\end{figure*}

\begin{figure*}
\begin{center}
\includegraphics[width=160mm]{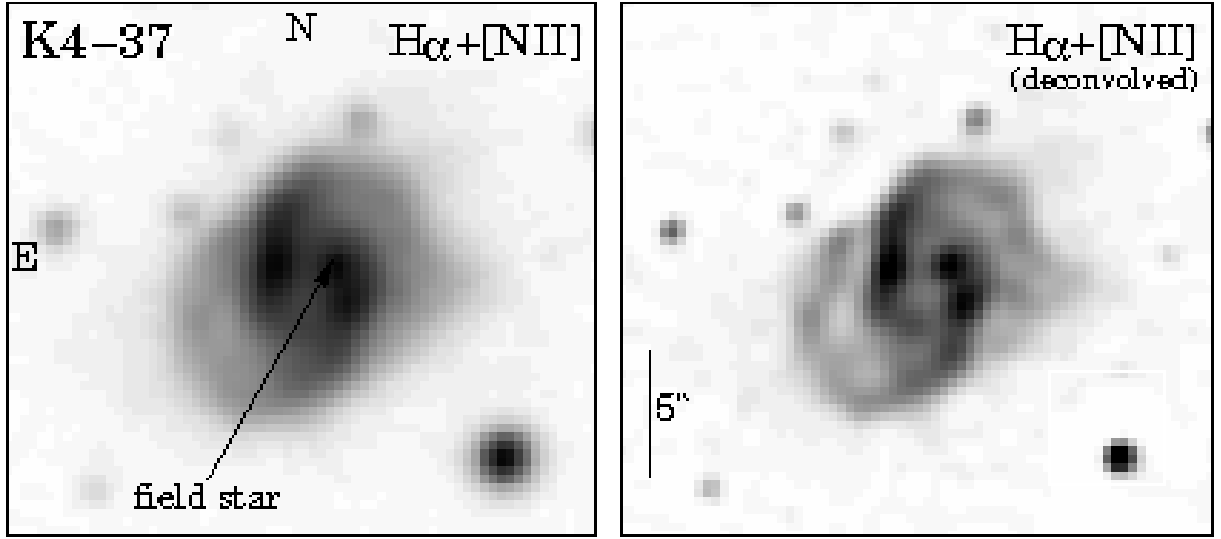}
\caption{Grey-scale reproductions of the 2014 original (left) and deconvolved
  (right) H$\alpha$+[N\,{\sc ii}] image of K\,4-37.  The grey levels are
  logarithmic. North is up, east to the left. A field star is arrowed in the
  H$\alpha$+[N\,{\sc ii}] original image (see caption of Fig.\,1). The size of the
  field shown in the two panels is 23$\times$21.5\,arcsec$^2$.}
\end{center}
\end{figure*}

\begin{figure}
\begin{center}
\includegraphics[width=80mm]{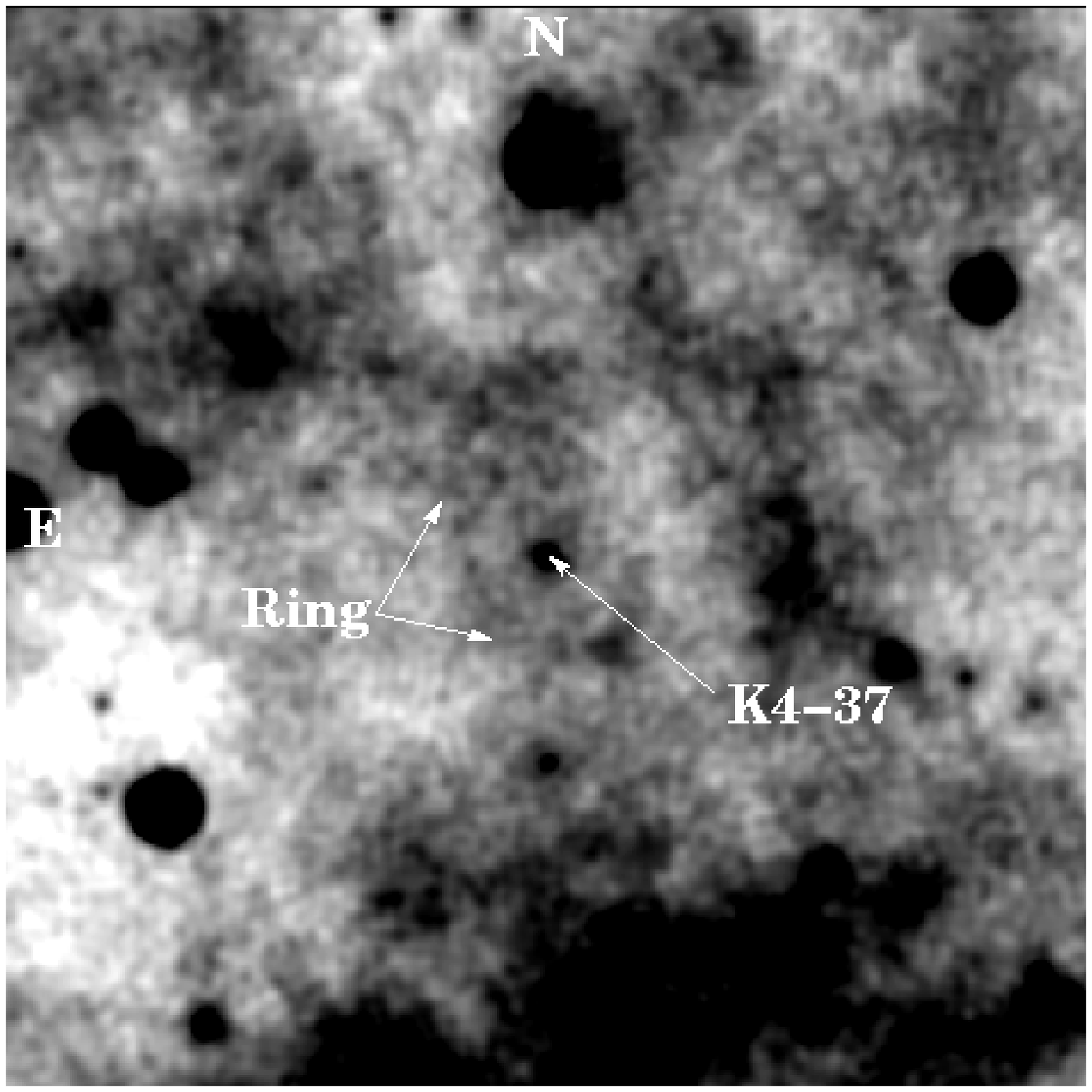}
\caption{Grey-scale reproduction of the WISE image at 22\,$\mu$m (W4) around K\,4-37
  (arrowed). Grey-levels are linear. North is up, east to the left, and the field of view is 
15$\times$15\,arcmin$^2$. The elliptical ring-like structure is arrowed.}
\end{center}
\end{figure}

It is well known that planetary nebulae (PNe) originate from Asymptotic Giant
Branch (AGB) and post-AGB stars with low- and intermediate-mass ($\la$ 
8--10\,M$_ {\odot}$) main-sequence progenitors once the central star becomes hot 
enough to photoionize the envelope ejected during the AGB phase. It is also well 
known that this process, when applied to single star evolution, can hardly explain
the dramatic transformation of a spherical AGB envelope into the varied and
complex morphologies observed in a noticeable number of PNe (e.g., Balick \& Frank 2002). 

During many years the formation of PNe has been understood
within the Interacting Stellar Wind model (ISW; Kwok, Purton \& Fitzgeral
1978; Balick 1987). This model has probed its ability to explain many of the features observed
in PNe (Balick \& Frank 2002, and references therein). However, multiple,
highly collimated, and axisymmetric structures in PNe have always presented
difficulties to be explained within the ISW scenario. Nowadays, in addition
to wind-wind interaction, it is recognized  that the formation of complex PNe 
requires the action of collimated outflows that distort an originally spherical AGB envelope 
(Sahai \& Trauger 1998). In this context, binary central stars were always though to be 
determining in the formation of complex PNe (e.g., Morris 1981; Soker \& Livio 1994). 
The importance of binary central stars in PN
formation is receiving strong support with the detection of new post-common
envelope binary central stars associated to complex PNe (Miszalski et
al. 2009; Miszalski 2012; Jones 2016; and references therein). Evolution
through a common-envelope provides many of basic ingredients though
to be involved in the formation of complex PNe, as accretion disks, highly 
collimated outflows, and magnetic fields (Garc\'{\i}a-Segura et al. 1999; Dennis et al. 2008; de Marco 2009; 
and references therein). 

Nowadays, about 3500 PNe are known in the Galaxy (Parker, Boji\v{c}i\'c \& 
Frew 2016). However, many of them still lack an analysis that allow us to place them 
within the evolutionary sequence and to carry out comparisons with other PNe, which 
could provide information to disetangle the possible key processes involved in 
the formation of the different kinds of PNe. Given the enormous variety of properties (e.g., morphology, kinematics, 
chemical abundances, physical conditions) exhibited by PNe, increasing the number of well 
analyzed objects may contribute to shed light on the processes involved in the AGB to PN transition.

K\,4-37 (PN\,G066.9+02.2; $\alpha$(2000.0) = $19^{\rm h}$ $51^{\rm m}$ 
$00\rlap.^{\rm s}6$; $\delta$(2000.0) = $+31^{\circ}$ 02$'$ 29$''$) is a PN that has never 
been analyzed in detail. It was discovered by Kohoutek
(1965) and its PN nature can be guessed from the emission line fluxes listed
in the Strasbourg ESO Catalog (Acker et al. 1992). According to this catalog,
the nebula presents very strong [N\,{\sc ii}] and [S\,{\sc ii}], moderate [O\,{\sc iii}], and
absent He\,{\sc ii}\,$\lambda$4686 emission lines, indicating a very low-excitation PN. 
Except for these data, information about the characteristics and properties of K\,4-37 is lacking 
in the published literature. The object was imaged by us many years ago in our programs to
study compact PNe (e.g., Miranda et al. 1997, 2000, 2001;
V\'azquez et al. 2002), and it showed a simple and less interesting bipolar
morphology. However, the acquisition and analysis of new data, 
and inspection of the WISE archive have revealed noticeable properties in this
PN. In this paper we present the data, analysis, and results that we have
obtained for K\,4-37.

\section{Observations and results}

\subsection{Optical imaging}

Narrow-band images of K\,4-37 were obtained during 1998 July 7-9 with CAFOS 
at the Calar Alto Observatory. The detector was LORAL 2k$\times$2k CCD with a plate scale of 
0.33\,arcsec\,pixel$^{-1}$. Three filters were used: H$\alpha$ (FWHM =
15\,{\AA}), [N\,{\sc ii}] (FWHM = 19\,{\AA}), and [O\,{\sc iii}] (FWHM =
25\,{\AA}). Exposure time was 600\,s for each filter. The
spatial resolution, determined by the FWHM of field stars, is 1.2\,arcsec in the 
H$\alpha$ and [N\,{\sc ii}] images and 3\,arcsec in the [O\,{\sc iii}] one. 

Additional narrow-band images were obtained also with CAFOS in 2014 June 4 
using an H$\alpha$+[N\,{\sc ii}] filter (FWHM = 50\,{\AA}) and an exposure
time of 3$\times$600\,s, and in 2016 May 3 using an [O\,{\sc iii}] filter
(FWHM = 25\,{\AA}) and an exposure time of 2$\times$900\,s. 
In 2014 and 2016 the detector was a SITe 2k$\times$2k CCD with a plate scale of
0.53\,arcsec\,pixel$^{-1}$. The FWHM of the field stars in the H$\alpha$+[N\,{\sc ii}] 
image is 0.9\,arcsec, lower than twice the pixel size, suggesting that the true spatial resolution of 
this image is better than 0.9\,arcsec. In the 2016 [O\,{\sc iii}] image, the spatial resolution 
is 1.4\,arcsec.

The images were cosmic-ray cleaned, bias subtracted and flat fielded
using standard routines within the {\sc midas} package. To gain a better view
of the morphology of K\,4-37, we have used the Richardson-Lucy algorithm in the
{\sc midas} package, to deconvolve the H$\alpha$+[N\,{\sc ii}] image. A well exposed field star 
in the image was used as point-spread-function, and iteration was stopped when artifacts appeared in 
the deconvolved image. After 30 iterations, we obtained a deconvolved H$\alpha$+[N\,{\sc ii}] image with 
a spatial resolution of $\sim$0.5\,arcsec.

\begin{figure*}
\begin{center}
\includegraphics[width=160mm]{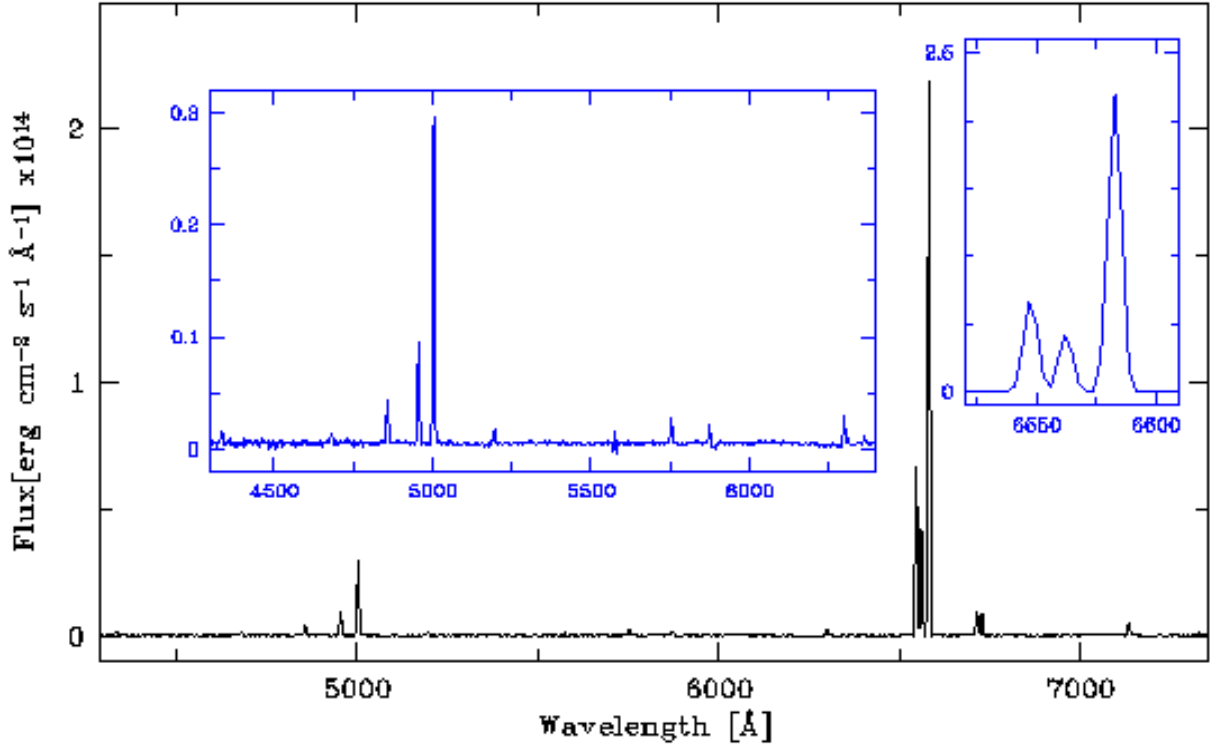}
\caption{Flux-calibrated spectrum of K\,4-37. The left inset shows the
  spectrum between 4300 and 6400 {\AA } and the right inset in the [N\,{\sc
    ii}]$\lambda$$\lambda$6548,6583 and H$\alpha$ region.[A color verison of the figure is available 
in the online version.]}
\end{center}
\end{figure*}

The 1998 H$\alpha$ and [N\,{\sc ii}] images together with the 2016 [O\,{\sc iii}] image are 
shown in Figure\,1. The original and deconvolved H$\alpha$+[N\,{\sc ii}] images are presented in Figure\,2. 
The nebula show a bipolar shape with a size of $\sim$12$\times$6.5\,arcsec$^2$ in 
H$\alpha$ and [N\,{\sc ii}], somewhat smaller in [O\,{\sc iii}], and the major
axis oriented at PA $\sim$$-$40$^{\circ}$. Nevertheless, faint emission can 
be detected in a $\sim$18$\times$11\,arcsec$^2$ area. The images show a point-symmetric 
intensity distribution in the bipolar lobes approximately along the north-south direction. A 
torus traces the equatorial plane of the nebula, which contains two bright knots located symmetrically 
with respect to the center of the object and oriented at PA $\sim$60$^{\circ}$. Due to the 
point-symmetric regions of the lobes, the axis of the torus seems to be oriented close to the 
east-west direction. However, the deconvolved H$\alpha$+[N\,{\sc ii}] image does show that its axis 
is oriented similarly to that of the bipolar lobes. The size of the torus is $\sim$4$\times$2\,arcsec$^2$ 
and some distortions can be recognized towards its southeastern side.

More details about the morphology of K\,4-37 can be recognized in the original and
deconvolved H$\alpha$+[N\,{\sc ii}] images (Fig.\,2). These
images show a noticeably limb-brightening and a knotty structure in the bipolar lobes. 
Moreover, the northwestern lobe is clearly distorted or 
broken towards the west where an elongated feature is observed up to $\sim$7.3\,arcsec from the center, 
which clearly deviates from the lobe geometry. An 
eastern counterpart cannot be recognized in the southeastern lobe (Fig.\,2)
but we note that the region along PA $\sim$90$^{\circ}$ protrudes and does not follow 
the lobe geometry. In fact, the southeastern lobe can be traced up to $\sim$5.7\,arcsec from
the center along PA $\sim$90$^{\circ}$ whereas it extends only $\sim$4\,arcsec 
at PA $\sim$180$^{\circ}$. The distorted regions subtend 
an angle of $\sim$30$^{\circ}$ as seen from the center. From the images in
Fig.\,2 two different axes are identified in K\,4-37: one
along PA $\sim$$-$40$^{\circ}$, defined by the main axis of the lobes 
and bright torus, and another one along PA $\sim$270$^{\circ}$ (or $\sim$90$^{\circ}$), defined by the
distortions of the bipolar lobes. Besides, the point-symmetric features suggest additional 
particular directions in the nebula.

The central star of K\,4-37 has never been identified and is neither observed in 
our narrow-band images. Central stars with $V$ or $B$ magnitudes $\la$ 17$^{\rm m}$ can be 
detected in (narrow-band) H$\alpha$ or [O\,{\sc iii}] images with moderate exposure times, 
which implies that the central 
star of K\,4-37 should be very faint. In an attempt to detect it, we obtained an image of 
K\,4-37 with CAFOS in 2016 May 3 using a Johnson B filter and an exposure time of 1200\,s, 
under seeing conditions of 
$\sim$1.5\,arcsec. No star can be recognized in our B image close to center of the object, 
where the central star could be expected, which would imply a lower limit of $\sim$20$^{\rm m}$ 
for its $B$ magnitude. On the other hand, it is well known that central stars may be displaced from 
the nebular center (e.g., Soker 1994; Tweedy \& Napiwotzki 1994; Soker, Rappaport \& Harpaz 1998; 
Chiotellis et al. 2016). We note that the images of K\,4-37 show a star at $\sim$1.7\,arcsec from 
nebular center (Figures\,1 and 2). Its spectrum indicates a K-type 
star (see Section\,2.3). A (somewhat speculative) alternative to the non detection of a star 
at the nebular center could be that the central star and the K-type star constitute a 
(non-resolved) binary system that has moved since the nebula was formed. If this was the case, 
the angular distance to the center and the age of the nebula ($\sim$10$^4$\,yr, Section\,3.2) 
would imply a proper motion of $\sim$0.17\,mas\,yr$^{-1}$ for the pair, corresponding to a 
tangential velocity of $\sim$11\,km\,s$^{-1}$ (at a distance of 14\,kpc, see Section\,3.2). If so, 
the large distorsion of the nebula towards the west, as compared with that towards the east, 
could be partially due to that motion. In any case, further efforts to detect the central star of 
K\,4-37 would be very valuable. 

\subsection{WISE imaging}

To complement our analysis, we have retrieved the images of K\,4-37 obtained with the Wide-field Infrared 
Survey Explorer (WISE) satellite in the four bands: W1 (3.4\,$\mu$m), W2 (4.6\,$\mu$m), W3 (12\,$\mu$m), 
and W4 (22\,$\mu$m), in a field of 15$\times$15\,arcmin$^2$ centered on the object. 

K\,4-37 itself is not resolved in the WISE images owing their relatively low spatial resolution 
(6.1--12\,arcsec) and the small size of the object. In addition, the field
star observed in the optical images (Fig.\,2) is bright at 3.4--22\,$\mu$m 
and this star and the nebula are indistinguishable in the WISE images. The more interesting result  
is observed in the W4 band a reproduction of which is shown in Figure\,3. K\,4-37 appears embedded 
in a complex nebular environment in which stands
out an elliptical ring-like structure surrounding the optical PN. Emission from the ring-like structure is also 
observed in the W3 band but more diffuse and fainter than in the W4 one, and is not observed in the W1 and W2 bands  
(not shown here). The elliptical ring-like structure (hereafter referred to as the 22\,$\mu$m
ring) is open (or very faint) towards the northwest, contains a bright spot towards the southwest, and presents 
a complex region towards the northeast where it seems to contact an 
ambient filament. The edge of the 22\,$\mu$m ring is relatively bright but emission is
also detected inside it. The complex northern region makes it
difficult to obtain accurately the orientation and major axis of this structure. Nevertheless, 
its minor axis appears oriented close to PA $\sim$$-$60$^{\circ}$ and is $\sim$2\,arcmin in size, 
its major axis is $\sim$2.9--3.6\,arcmin in size as measured at PA
30$^{\circ}$ (in the following we will adopt a major axis of 3.2\,arcmin), while a
size of $\sim$2.4\,arcmin is measured at PAs 75$^{\circ}$ and
165$^{\circ}$. From the image in Fig.\,3, an association, and not a
superposition chance, between the 22\,$\mu$m  ring and the optical nebula is strongly
suggested because the 22\,$\mu$m ring, the only closed structure observed in the field, is 
well centered on K\,4-37. Nevertheless, the optical nebula 
seems to be slightly displaced towards the southeast with respect to the possible 
center of the 22\,$\mu$m ring, which is difficult to define because of its complex
northern region. The large difference in sizes between the optical nebula and
the 22\,$\mu$m ring strongly suggests that the latter should correspond to a much older ejection
than that involved in the formation of the main nebula. As we will discuss
below (Section\,3.3), the 22\,$\mu$m ring probably represents mass ejected from the AGB 
progenitor of K\,4-37. We note that the detection of
this structure at 22\,$\mu$m suggests dust as origin of its emission.

\subsection{Intermediate-resolution long-slit spectroscopy}

Intermediate-resolution, long slit-spectra of K\,4-37 were obtained on 2008 June 8 with
the Boller \& Chivens spectrograph mounted on the 2.1\,m telescope at the San
Pedro Martir Observatory (OAN-SPM)\footnote{The Observatorio Astron\'omico Nacional at the Sierra de 
San Pedro M\'artir (OAN-SPM) is operated by the Instituto de Astronom\'{\i}a 
of the Universidad Nacional Aut\'onoma de M\' exico}, using the SITe3 CCD (plate scale
24\,$\mu$m~pix$^{-1}$), with a 1k$\times$1k pixel array as a detector. We used a
400\,lines\,mm$^{-1}$ dispersion grating along with a 2.5\,arcsec slit
width, giving a spectral resolution (FWHM) of $\sim$6.5\,{\AA} and covering
the 4300--7400\,{\AA} spectral range. The slit was centered on the nebula and oriented at PA
90$^{\circ}$. Four spectra were obtained with an exposure time of 1800\,s each. Spectra reduction was carried 
out following standard procedures in XVISTA\footnote{XVISTA was originally
  developed as Lick Observatory Vista. It is currently mantained by Jon
  Holtzman at New Mexico State University and is available at
  http://ganymede.nmsu.edu/holtz/xvista.}. In particular, the median of
the four spectra was first derived and then it was subtracted from each individual spectrum to 
detect and eliminate cosmic rays manually. Finally, the cleaned spectra were
added to get a final spectrum with a total exposure time of
7200\,s. Spectrophotometric standard stars were observed in the same night for
flux calibration purposes. Seeing was $\sim$2.5\,arcsec during the observations. 

The field star overimposed on the nebula (Figs.\,1 and 2) is
covered by the slit. The spectrum indicates a K-type for this star. We have
fitted the continuum of this star and subtracted it from the integrated nebular spectrum. 
The contribution of the stellar absorption lines to the Balmer nebular emission lines is 
difficult to evaluate because a more precise spectral type cannot be deduced for
the field star. Nevertheless, from other observed absorption lines in the
stellar spectrum, we estimate that this contribution should be
lower than $\sim$2\% and, therefore, it is not critical in the analysis of the 
nebular emission spectrum. 

Figure\,4 shows the integrated spectrum of K\,4-37 
and Table\,1 lists the emission line intensities, their errors, and several parameters, 
as determined with the programm {\sc anneb}, a description of which can be found in Olgu\'{\i}n et al. (2011). 
Briefly, we used the extinction law of Seaton (1979) and assume case B recombination to
start an iterative process until the values of the logarithmic extinction
coefficiente $c$(H$\beta$), electron temperatures $T$$_{\rm e}$$[\ion{N}{ii}]$ and
$T$$_{\rm e}$$[\ion{O}{iii}]$ obtained from the [N\,{\sc ii}] and [O\,{\sc iii}] emission line ratios,
respectively, and electron density, $N$$_{\rm e}$$[\ion{S}{ii}]$ obtained from the [S\,{\sc
  ii}]$\lambda$$\lambda$6716,6731 emission lines, converge to their final values that are listed 
in Table\,1.

Figure\,4 and Table\,1 show the extreme low-excitation of
K\,4-37 with $[\ion{N}{ii}]$/H$\alpha$ and $[\ion{S}{ii}]$/H$\alpha$ line
intensity ratios of $\sim$6.6 and $\sim$0.4, respectively, and
prominent $[\ion{N}{i}]$ and $[\ion{O}{i}]$ emission lines as compared
with that usually observed in PNe (see Section\,3.1). The $[\ion{O}{iii}]$/H$\beta$
line intensity ratio presents a moderate value of $\sim$7.7. Noticeably, the
$\ion{He}{ii}$\,$\lambda$4686 emission line is also detected in our spectrum and
indeed with a relatively strong intensity of about a third of H$\beta$, suggesting a relatively 
hot central star. Finally, the electron temperatures are quite typical of PNe, while the electron 
density is very low (see Table\,1), indicating a very evolved PNe. 

Ionic and elemental abundances were derived using the values of $N$$_{\rm e}$$[\ion{S}{ii}]$
and $T$$_{\rm e}$$[\ion{N}{ii}]$ in Table\,1, and are listed
in Table\,2. The icf method of Kingsburgh \& Barlow (1984) has been used as implemented in {\sc
  anneb}. The high He (He/H $\sim$0.145) and N (N/H $\sim$6.3$\times$10$^{-4}$)
abundances make K\,4-37 a typical type\,I PN
(Peimbert 1990). Sulfur and argon abundances are in the range of the values observed in
PN and type\,I PNe (e.g., Kingsburgh \& Barlow 1994). What makes K\,4-37
particularly interesting is the high value of the N/O abundance ratio of
$\sim$2.32 that is among the highest ones observed in PNe (see, e.g., Pottasch, Beintema \& 
Feibelman 2000; Akras et al. 2016 and references therein) and indicates an extreme nitrogen 
enrichment. The high He and N abundances, and N/O abundance ratio imply an intermediate mass
progenitor. We have compared the values in Table\,2 with the models for stellar yields 
by Karakas (2010) to estimate an initial mass of 4--6\,M$_{\odot}$ for the main-sequence 
progenitor of K\,4-37. 

\begin{table}
\centering  
\caption{Dereddened emission line intensities in K\,4-37 in units of 
$I_{\rm H\beta}$ = 100.0}                           
\begin{tabular}{lcrr}
\hline
Ion                       &  $\lambda$ & $f$($\lambda$) & $I$($\lambda$)  \\
\hline
H$\gamma$                 & 4340   &   0.129 &    64.5  $\pm$ 5.9 \\ 
$[\ion{O}{iii}]$          & 4363   &   0.124 &    12.2  $\pm$  3.7 \\ 
\ion{He}{ii}              & 4686   &   0.042 &    29.9   $\pm$    2.8 \\
H$\beta$                  & 4861   &   0.000 &   100.0    $\pm$  2.3 \\  
$[\ion{O}{iii}]$          & 4959   &  --0.024 &   194.5 $\pm$    3.4 \\ 
$[\ion{O}{iii}]$          & 5007   &  --0.035 &   572.5  $\pm$    9.5 \\ 
$[\ion{N}{i}]$            & 5199   &  --0.075 &    26.0    $\pm$  1.1 \\ 
$[\ion{N}{ii}]$           & 5755   &  --0.192 &    24.0   $\pm$    0.7 \\ 
\ion{He}{i}               & 5876   &  --0.216 &    15.3  $\pm$  0.6 \\ 
$[\ion{O}{i}]$            & 6300   &  --0.285 &    20.2   $\pm$  0.6 \\ 
$[\ion{S}{iii}]$          & 6312   &  --0.287 &     5.5   $\pm$    0.4 \\ 
$[\ion{O}{i}]$            & 6363   &  --0.294 &     5.5   $\pm$    0.3 \\ 
$[\ion{N}{ii}]$           & 6548   &  --0.321 &   465.9   $\pm$   13.0 \\ 
H$\alpha$                 & 6563   &  --0.323 &   283.9   $\pm$   8.0 \\ 
$[\ion{N}{ii}]$           & 6583   &  --0.326 &  1414.8   $\pm$   39.8 \\ 
\ion{He}{i}               & 6678   &  --0.339 &     5.2    $\pm$  0.3 \\ 
$[\ion{S}{ii}]$           & 6716   &  --0.343 &    60.0    $\pm$  1.8 \\ 
$[\ion{S}{ii}]$           & 6731   &  --0.345 &    53.6    $\pm$  1.6 \\ 
\ion{He}{i}               & 7065   &  --0.383 &     4.8    $\pm$  0.3 \\ 
$[\ion{Ar}{iii}]$         & 7136   &  --0.391 &    25.7   $\pm$  0.9 \\ 
$[\ion{O}{ii}]$           & 7320   &  --0.411 &     4.9   $\pm$    0.3 \\ 
$[\ion{O}{ii}]$           & 7330   &  --0.412 &     5.1   $\pm$    0.3 \\
\hline
log\,$F$(H$\beta$)          &        &          &  --16.23 $\pm$ 0.03 \\
(erg\,cm$^{-2}$\,s$^{1}$)  &        &          &                      \\ 
$c$(H$\beta$)               &        &          &    1.71 $\pm$ 0.03 \\
$T$$_{\rm e}$$[\ion{N}{ii}]$ (K)        &        &          & 11000 $\pm$ 300 \\
$T$$_{\rm e}$$[\ion{O}{iii}]$ (K)       &        &          & 15800 $\pm$ 2450 \\
$N$$_{\rm e}$$[\ion{S}{ii}]$ (cm$^{-3}$)       &        &          & 370 $\pm$ 120 \\ 
\hline
\end{tabular}
\end{table}

\begin{table}
 \centering  
\caption{Ionic and elemental abundances by number in K\,4-37}                           
\begin{tabular}{lr}
\hline
Ion                           & Abundance \\
\hline
He$^{+}$ ($\times$$10^{2}$) &  12.060 $\pm$   0.397 \\ 
He$^{2+}$ ($\times$$10^{2}$) &   2.42 $\pm$   0.22 \\ 
O$^{0}$  ($\times$$10^{4}$) &   0.26 $\pm$   0.01 \\ 
O$^{+}$  ($\times$$10^{4}$) &   0.93 $\pm$   0.04 \\ 
O$^{2+}$  ($\times$$10^{4}$) &   1.48 $\pm$   0.02 \\ 
N$^{0}$  ($\times$$10^{5}$) &   2.61 $\pm$   0.11 \\ 
N$^{+}$  ($\times$$10^{4}$) &   2.15 $\pm$   0.04 \\ 
S$^{+}$  ($\times$$10^{6}$) &   2.21 $\pm$   0.05 \\ 
S$^{2+}$  ($\times$$10^{6}$) &   8.93 $\pm$   0.64 \\ 
Ar$^{2+}$ ($\times$$10^{6}$) &   1.90 $\pm$   0.06 \\ 
\hline
Element                      &  Abundance   \\
\hline
He/H                       & 0.145 $\pm$ 0.005 \\ 
O/H    ($\times$$10^{4}$)  & 2.72 $\pm$ 0.06 \\ 
N/H    ($\times$$10^{4}$)  & 6.31 $\pm$ 0.36 \\ 
Ar/H    ($\times$$10^{6}$)  & 3.55 $\pm$ 0.79 \\ 
S/H     ($\times$$10^{5}$)  & 1.25 $\pm$ 0.07 \\ 
\hline
\end{tabular}
\end{table}

\subsection{High-resolution long-slit spectroscopy}

High-dispersion optical spectra were obtained with the Manchester Echelle
Spectrometer (Meaburn et al. 2003) and the 2.1\,m (f/7.5) telescope at the 
OAN-SPM Observatory during 2004 July 29$-$30. A SITe CCD with 1k$\times$1k pixels was used as detector. 
The slit length is 6.5\,arcmin and its width was set to 150\,$\mu$m (2\,arcsec).
A 2$\times$2 binning was used, leading to a spatial scale of 0.66\,arcsec\,pixel$^{-1}$ 
and a spectral scale of 0.1\,{\AA}\,pixel$^{-1}$. This spectrograph has no cross
dispersion, consequently, a $\Delta\lambda=90$\,{\AA} bandwidth filter was used 
to isolate the 87$^{\rm th}$ order covering the spectral range that includes
the H$\alpha$ and [\ion{N}{ii}]$\lambda$$\lambda$6548,6583 emission lines. Spectra were
obtained with the slit centered across the center of the nebula and oriented
at PAs $-40\degr$ and $+90\degr$. Exposure time for
each spectrum was 1200\,s. The spectra were wavelength calibrated with a Th-Ar
arc lamp to an accuracy of $\pm1$\,km\,s$^{-1}$. The FWHM of the arc lamp
emission lines was measured to be $\simeq12$\,km\,s$^{-1}$ that corresponds to the achived spectral
resolution. Seeing was $\sim$2\,arcsec during the observations. Reduction of the
spectra was carried out with standard routines in the {\sc iraf} package.

\begin{figure*}
\begin{center}
\includegraphics[width=140mm]{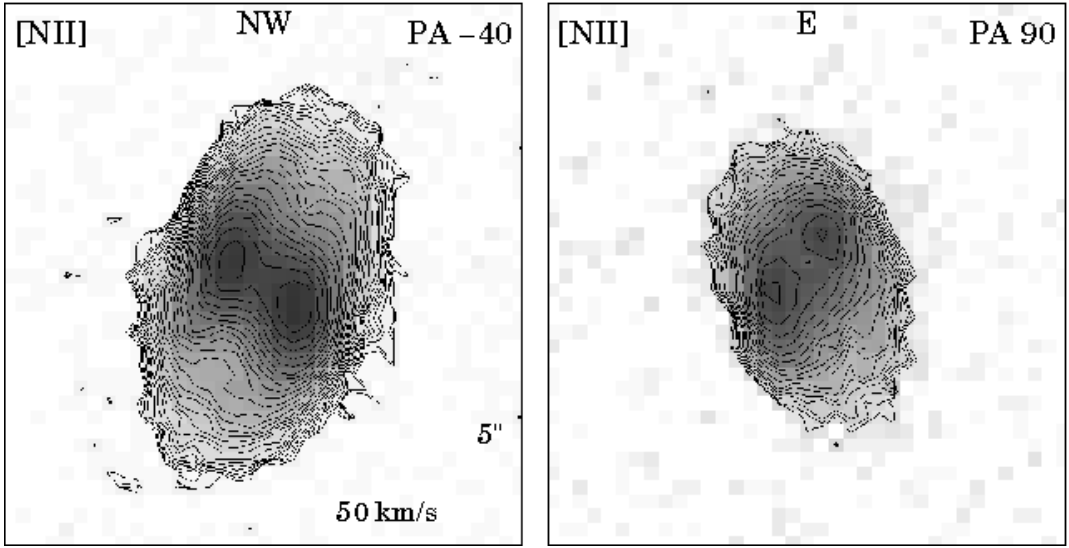}
\caption{Grey-scale and contours position-velocity maps of the [N\,{\sc
  ii}]$\lambda$6583 emission line at the two observed PAs (upper right). The grey levels and
contours are logarithmic. Spatial and velocity scales are identical in both panels and indicated in the left one.}
\end{center}
\end{figure*}

\begin{figure*}
\begin{center}
\includegraphics[width=160mm]{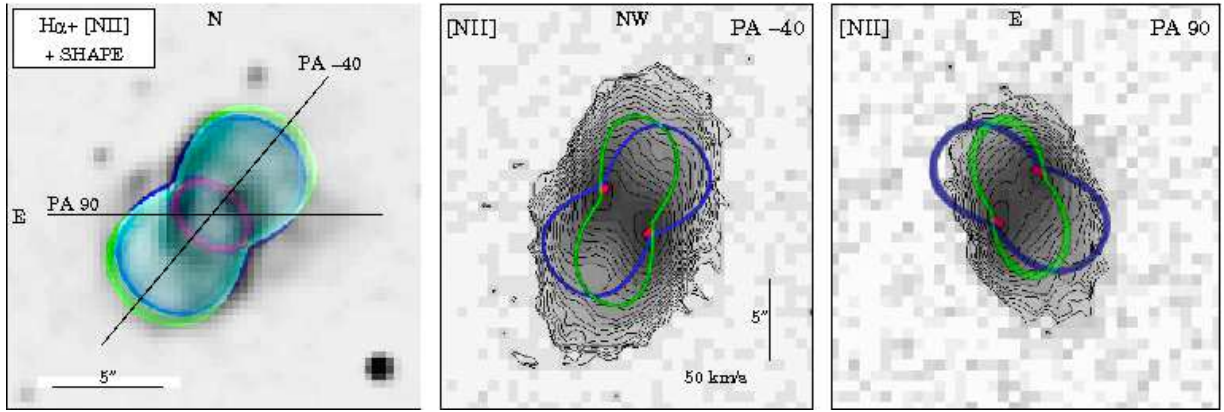}
\caption{Grey-scale H$\alpha$+[N\,{\sc
  ii}] deconvolved image (left), and grey-scale and contours position-velocity
(PV) maps of the [N\,{\sc ii}]$\lambda$6584 emission line (middle and right). The grey levels and
contours are logarithmic. The orientation of the long slits are plotted on the image (slit width not
to scale), and the PV maps contain information about the orientation, and spatial and
  velocity scales the latter two being identical in the middle and right panels. Overimposed on the 
panels are the results of {\sc shape}: red color corresponds to the equatorial torus, blue color to the
  bipolar lobes assuming homologous
  expansion velocity, and green color to the bipolar lobes assuming non-homologous expansion velocity and an
  inclination angle for the bipolar lobes different from that of the torus (see
  text for details). [A colour version of this figure is available in the online version.] }
\end{center}
\end{figure*}

Figure\,5 presents position-velocity (PV) maps of the [N\,{\sc
  ii}]$\lambda$6583 emission line at the two observed PAs. From these high-resolution spectra,
the systemic velocity of K\,4-37 was measured by co-adding 2.4\,arcsec around the central
position of the emission line features observed at each PA, and finding the central
radial velocity of the resulting line profile. The estimates for the systemic
velocity in both spectra and from the H$\alpha$ and [N\,{\sc
  ii}]$\lambda$$\lambda$6548,6583 emission
lines are consistent with each other after heliocentric and 
LSR corrections, and its mean value is $V_{\rm HEL}$ = $-$23.7 $\pm$
1.5\,km\,s$^{-1}$ ($V_{\rm LSR}$ = $-$5.3 $\pm$ 1.5\,km\,s$^{-1}$). 

The [N\,{\sc ii}] emission feature in the PV maps (Fig.\,5) is dominated by two
intensity maxima that correspond to the equatorial ring. The PV map at PA
$-$40$^{\circ}$ shows two velocity components at each spatial position, as expected from a bipolar
shell. The velocity splitting is small and varies between $\sim$22 and $\sim$30\,km\,s$^{-1}$. 
At PA 90$^{\circ}$ velocity splitting is clearly observed around the center of the line 
feature but is difficult to recognize at larger angular distances. The long-slit
spectrum at PA 90$^{\circ}$ covers the distorted region towards the west (Fig.\,2). 
In fact, the [N\,{\sc ii}] emission feature in the PV map at PA 90$^{\circ}$ is much more 
extended towards the west than towards the east, in consonance
with the images (Fig.\,2). If the weastern nebular regions have been distorted, as the images show, 
its kinematics may have been distorted, too, which may explain the lack of velocity splitting, as 
expected from a bipolar shell. 

To extract more information from the PV maps, we used the tool {\sc shape} (Steffen et al. 2011) 
to find the morphokinematic parameters that are able to reproduce
simultaneously the observed morphology and PV maps. It should be emphasized that we did not try 
to reproduce with {\sc shape} all the observed features in the PV maps and image in detail as, e.g.,  
changes of intensity (point-symmetric regions, bright knots), (spatial and velocity) widths of the features, 
and distortions of the structures. 
Rather, we try to reproduce the variation of the radial velocity (two components) as a 
function of the spatial position, as observed in the PV maps, and the basic shape (projected contour) 
of the bipolar lobes and torus, as observed in the image. In Figure\,6 we show 
the deconvolved H$\alpha$+[N\,{\sc ii}] image and [N\,{\sc ii}] PV maps of K\,4-37
together with the results of {\sc shape} assuming a thin bipolar shell with a homologous expansion velocity law, 
i.e., expansion velocity ($V$$_{\rm exp}$) proportional to the radius (r) at each point of the nebula 
[$V$$_{\rm exp}$($r$) = $K$$\times$$r$] and narrow slits (1\,arcsec) at PAs --40$^{\circ}$ and 90$^{\circ}$ 
(but see below).

The equatorial torus is quite well reproduced with a circular structure of 1.9\,arcsec in radius, 
expansion velocity of 15.3\,km\,s$^{-1}$, and a tilt for its axis of 45$^{\circ}$
with respect to the plane of the sky (red colour in Figure\,6). These parameters 
imply $K$ $\sim$8.0\,km\,s$^{-1}$\,arcsec$^{-1}$. When we applied 
this law to the bipolar lobes, we found that their kinematics 
cannot be reproduced satisfactorily, even though their morphology can be reproduced
acceptably (blue colour in Figure\,6). It is clear that the kinematics 
of the bipolar lobes predicted assuming homologous expansion results to be very different 
from the observed one. Firstly, we have considered whether a slit width of 2\,arcsec 
(as we used for the long-slit spectra, see above) could have an effect on the resulting 
shape of the blue lines in Figure\,6. However, the use of a 2\,arcsec slit produces virtually 
the same results as shown in Figure\,6, except that the blue lines become slightly thicker 
inwards as a result of including nebular regions at both sides of the main axis, which 
have somewhat lower radial velocities than along the main axis. Other effects, as, e.g., 
smoothing due to seeing, may also increase the thickness of the blue lines but 
the {\it apparent tilt} of these lines in the PV maps {\it remains unchanged}. Then, we have modified 
(one by one and simultaneously) several parameters involved 
in the analysis (e.g., size of the polar axis, polar expansion velocity, inclination 
angle of the polar axis, in all cases also with a 2\,arcsec slit width) and our conclusion is that reproducing 
the morphokinematics of the bipolar lobes requires that the main axis of the lobes is not perpendicular to the
plane of the torus, and, in addition, that a non-homologous expansion law should be considered. In Fig.\,6 (green colour) 
we also show the morphology and kinematics expected if the main axis of the lobes is tilted by
20$^{\circ}$ with respect to the plane of the sky, the semi-major axis
is 5.3\,arcsec, and the polar expansion velocity is 34\,km\,s$^{-1}$. These parameters 
provide a much more acceptable reproduction of the kinematics observed in the PV maps 
and keep an acceptable representation of the morphology of the lobes as well. They also  
imply $K$ $\sim$6.4\,km\,s$^{-1}$\,arcsec$^{-1}$ for the bipolar lobes, that is 
sensibly different from that obtained for the equatorial torus. We note that the western region of the PV map at 
PA +90$^{\circ}$ is difficult to reproduce (Fig.\,6), and emission is
observed at higher redshifted radial velocities than the predicted ones. This
suggests that the distorted region could be more tilted than the axis of the
lobes (i.e., $>$20$^{\circ}$ with respect to the plane of the sky), 
but a more precise determination would require spectra at higher spectral and spatial resolution. 
We have also checked the range of tilt angles capable to provide a reasonable reproduction 
of the observed morphokinematics and found that the deduced tilt angles are correct within 
$\sim$$\pm$3$^{\circ}$ for the axis of the bipolar lobes and
$\sim$$\pm$$1\rlap.^{\circ}5$ for the axis of the torus. 
These uncertainties are much smaller than the difference in tilt angles ($\sim$25$^{\circ}$), supporting that the
difference is real. Finally, we note that the expansion velocities obtained in
K\,4-37 ($\sim$15--35\,km\,s$^{-1}$), are noticeable lower than those observed in 
other bipolar PNe. 

The morphokinematical analysis has disclosed the presence of two different axes where 
only one is observed in the images as a the main bipolar axis. By chance, due to our line of sight, 
the axes of the torus and bipolar lobes appear aligned 
on the plane of the sky at about the same PA ($\sim$--40$^{\circ}$), but they present different tilts 
with respect to the observer (see also Section\,3.4 for the case of NGC\,6309).  
These two axes add to the axes of the distortions of the bipolar lobes and the orientation 
of the point-symmetric regions, which are all different from each other. According to these results, 
K\,4-37 may be better classified as a multi-axis PN rather than among bipolar PNe with a single axis. 

The non-homologous expansion found in K\,4-37 is worth mentioning. The idea
that PNe expand in an homologous manner has prevailed during long time mainly
due to the fact that a simple homologous expansion velocity law is able to 
reproduce the observed PV maps of many PNe in a quite satisfactory
manner. Nevertheless, PNe are known in which such a law is unable to reproduce
the observed PV maps (e.g., Miranda et al. 1999; Steffen, Garc\'{\i}a-Segura \& Koning
2009; Steffen et al. 2013; Berm\'udez 2015). In these cases, the action of collimated outflows on
the shell and interaction between the stellar wind and an
inhomogeneous shell have been invoked to explain the non-homologous expansion (see
references above). In this context, the formation of the bipolar lobes of
K\,4-37 seems to require a focussed wind/outflow along a direction that
is not perpendicular to the equatorial bright torus. In PNe, these kind of outflows are usually 
related to binary central stars. The possibility that at least in some PNe, non-homologous expansion 
could be suggesting a binary central star should be investigated.

From the morphokinematic parameters we obtain kinematical ages of $\sim$590$\times$D[kpc]\,yr 
and $\sim$740$\times$D[kpc]\,yr for the torus and bipolar lobes, respectively, that seem  
to imply that the formation of the bipolar lobes has preceded that of the torus. 
However, given the uncertainties involved in the analysis, we will 
consider a mean value of 665$\times$D[kpc]\,yr for the kinematical age of
K\,4-37. As of the 22\,$\mu$m ring, if we assume a constant expansion velocity of
15\,km\,s$^{-1}$ (typical of AGB envelopes), its kinematical age is
$\sim$3.1$\times$10$^4$$\times$D[kpc]\,yr, much larger than that of the
bipolar nebula.

\section{Discussion}


\subsection{The evolutionary status of K\,4-37}

If the global properties derived in the previous sections are considered, K\,4-37 seems to find 
its right place among the group of highly evolved PNe recently analyzed 
by Akras et al. (2016, hereafter AK16). Tables\,3 and 4 in AK16 and
Tables\,1 and 2 in this work show the common 
properties. These PNe, including K\,4-37, show very strong [N\,{\sc ii}]
and [S\,{\sc ii}] emissions that place them in the lower-left region of
the diagram log(H$\alpha$+[N\,{\sc ii}]) versus log(H$\alpha$+[S\,{\sc
  ii}]) (see Fig.\,9 in AK16). AK16 call the attention to the presence of unusually strong 
[N\,{\sc i}], [O\,{\sc i}] emission lines in their sample of highly evolved PNe, which suggest   
a contribution of shock excitation (see also Akras \& Gon\c calves 2016). K\,4-37 
presents [N\,{\sc i}] and [O\,{\sc i}] to H$\beta$ line intensity ratios comparable to those 
in the sample by AK16, and, therefore, shocks are probably contributing to the 
line emission. These PNe also present high He and N abundances, and high N/O abundance ratios, indicating 
progenitors of $\sim$5\,M$_{\odot}$, and very low electron densities (AK16; see Section\,2.3). To these common
characteristics, we add the presence of a relatively strong 
He\,{\sc ii}$\lambda$4686 nebular emission (He\,{\sc ii}$\lambda$4686/H$\beta$ $\sim$0.2--0.5; Table\,1; AK16 and 
references therein). This is probably related to the high effective temperature 
of their central stars (AK16). Although the central star of K\,4-37 has not
been detected, its pressumable faintness and relatively strong He\,{\sc ii}$\lambda$4686 nebular emisssion 
suggest that it could be of low luminosity and high effective temperature, as those in the sample 
analyzed by AK16.

\subsection{The distance of K\,4-37}

We have obtained a kinematical age of 665$\times$D[kpc]\,yr for K\,4-37 that depends on its 
distance (Section\,2.4). Although the distance is unknown, the resulting kinematical age should be 
compatible with K\,4-37 being a highly evolved PN. If, as a first approximation, we assume a kinematical 
age of $>$10000\,yr as representative of a highly evolved PN (see AK16), the distance to K\,4-37 
is $>$15\,kpc. A more confident distance can be obtained by using the recent surface 
brightness--radius relationship by Frew, Parker \& Boji\v{c}i\'c (2016), 
the observed H$\alpha$ flux and $c$(H$\beta$) (Table\,2), and a mean nebular radius 
of $\sim$6\,arcsec. With these values, the distance is $\sim$14\,kpc and, consequently, 
the kinematical age is $\sim$9300\,yr, compatible with a highly evolved PN. 

For 14\,kpc, the size of the nebula is $\sim$0.72$\times$0.26\,pc$^2$, very much smaller than 
the size of the 22\,$\mu$m ring that amounts $\sim$13$\times$8\,pc$^2$ and whose
kinematical age is of $\sim$4.3$\times$10$^5$\,yr. Taking into account the age of the nebula and 
that a 4--6\,M$_{\odot}$ star traverses the post-AGB phase in $\la$100\,yr (Bl\"ocker 1995), the 
kinematical age of the 22\,$\mu$m ring places its formation in the AGB 
phase of the progenitor of K\,4-37.

Finally, the low expansion velocities in K\,4-37 suggest that the shell might have been decelerated in
the course of the evolution. If this was the case, the ``true'' kinematical age of K\,4-37 would be $<$665$\times$D[kpc]\,yr, 
and 14\,kpc could be a lower limit to the distance to make compatible the kinematical age of K\,4-37 
with its evolved nature. In any case, it seems to 
be clear that K\,4-37 should be located at a relatively large distance.

\subsection{Mass ejection history in K\,4-37}

\begin{table*}
\centering  
\caption{Chemical abundances [12+log[$n$(X)/$n$(H)] in K\,4-37 and NGC\,6309.}                           
\begin{tabular}{lccccccl}
\hline
PN      & He & N & O & S & Ar & N/O & Reference \\
\hline
K\,4-37   & 11.161 & 8.80 & 8.43 & 7.10 & 6.55 & 2.32 & This work  \\ 
NGC\,6309 & 11.061 & 7.94 & 8.73 & 6.79 & 6.47 & 0.15 & V\'azquez et al. 2008 \\
 \hline
\end{tabular}
\end{table*}

Our data analysis has shown that K\,4-37 should be classified as a multi-axis
PNe. In addition, a very large elliptical structure has been found
at 22\,$\mu$m surrounded the optical nebula, which probably corresponds to mass ejected in the AGB phase 
of the progenitor of K\,4-37. These findings provide information about the formation history of K\,4-37 
 
According to the properties described in the previous sections, the 22\,$\mu$m ring 
probably is a detached shell related to mass ejected during the AGB phase. 
The emission detected inside the ring indicates that it is filled with material
while the bright edge traces the site of interaction between the AGB ejections and the ISM 
(e.g., Villaver, Garc\'{\i}a-Segura \& Manchado 2002; Sch\"oier, Lindqvist \& Olofsson 2005; Mattsson, 
H\"ofner \& Herwig 2007). Detached shells are observed in many AGB and post-AGB stars (e.g, Speck, Meixner 
\& Knapp 2000; Cox et al. 2012; and references therein). To the best of our knowledge, 
the 22\,$\mu$m ring in K\,4-37 is the largest one of this kind of shells, only comparable to that 
around R CrB (size $\sim$8.7\,pc, Gillett et al. 1997), and much larger than those around AGB stars 
(sizes $\sim$0.04--0.8\,pc, e.g., Olofsson et al. 2000; Cox et al. 2012), and post-AGB stars as, e.g., AFGL\,2688 and
CRL\,618 with sizes of $\sim$4 and $\sim$6.6\,pc, respectively (Speck et al. 2000). Although the evolution of
the these shells may be complex (e.g., Villaver et al. 2002), detached shells around AGB and post-AGB stars 
could be expected to grow noticeably when the (possible) associated future PNe reach a highly evolved phase. 

The elliptical morphology of the 22\,$\mu$m ring contradicts the
spherical mass ejection that is expected to dominate during the AGB phase
(e.g., Olofsson et al. 2000; Kerschbaum et al. 2010). It is interesting to
speculate about the origin of this morphology. Deformation of an
original spherical AGB shell by the own interaction with the ISM can be ruled out
because of the very peculiar ISM structure required to create a well defined
elliptical ring. The 22\,$\mu$m ring has some resemblance with 
the ``eye-type'' shells observed around some AGB stars (Cox et al. 2012). ``Eye-type'' shells 
have been modeled as deformation of an AGB spherical shell by the ISM
magnetic field (van Marle, Cox \& Decin 2015) and a similar model could be applicable to 
the 22\,$\mu$m ring. Alternatively, a companion could flatten the (spherical) envelope of an AGB star 
(Mauron, Huggins \& Cheung 2013) resulting in an oblated shell. Molecular line observations of the
22\,$\mu$m ring are required to study its kinematics, and to derive its 3D structure and 
properties. 

The formation of K\,4-37 itself fits well into the usual scenario in which 
several bipolar outflows have shaped the nebula (Sahai \& Trauger 1998). As
observed in many PNe, the bipolar outflows should have been ejected at different directions 
and their collimation degree has also changed; a poor collimated outflow seems to be 
involved in the formation of the bipolar lobes while more focused outflows could be
related to the distortions of the lobes and point-symmetric regions. Our data
do not allow us to stablish a possible time sequence in the generation of the
various outflows, although the bipolar lobes may be ascribed to a major
ejection event.

\subsection{Comparison of K\,4-37 and NGC\,6309}
 
Among multi-axis PNe, we found striking morphological similarities between
K\,4-37 and NGC\,6309, a PN with a bright equatorial torus and two
pairs of bipolar lobes at different directions (V\'azquez et al. 2008; Rubio et al. 2015). Knotty and
point-symmetric structures can be recognized in the lobes as well as
distortions in one of the pairs of NGC\,6309, characteristics that are
present in K\,4-37 as well. Interestingly, the two pairs of
bipolar lobes and the torus of NGC\,6309 seem to be tilted with respect to the observer by
the same amount (V\'azquez et al. 2008; Rubio et al. 2015). This 
implies that if NGC\,6309 was observed from a line of sight perpendicular to
the current one, it  would appear as a ``simple'' bipolar PN, with a pair of
bipolar lobes and an equatorial torus sharing a single main axis, and resembling the images of 
K\,4-37; in this case, only a morphokinematical analysis of NGC\,6309 would be able to 
reveal the existence of two pairs of lobes and different orientations of the structures. Moreover, 
a spherical halo surrounds NGC\,6309 and evidence exists for a large shell with a size of $\sim$1.5\,pc and
kinematical age of $\sim$1.5$\times$10$^5$\,yr (Rubio et al. 2015). This comparison 
strongly suggests that the central stars of
K\,4-37 and NGC\,6309 share a similar mass ejection history. In particular, the specific processes 
involved in the shaping of the multi-axis structure should have been 
very similar in both PNe.

Some differences can be noticed between K\,4-37 and NGC\,6309. In particular,
the electron density in NGC\,6309 (1400-4000\,cm$^{-3}$, V\'azquez et
al. 2008) does not indicate a particularly young or a highly evolved PN, in
consonance with its kinematical age of $\sim$4000\,yr (V\'azquez et al. 2008;
Rubio et al. 2015) that suggests a moderately young/evolved PN. Probably
K\,4-37 and NGC\,6309 are in a different evolutionary stage 
within PN evolution, being K\,4-37 in a more advanced one. 

However, the most interesting difference between K\,4-37 and NGC\,6309 is
observed in their chemical abundances that are listed in Table\,3. Although He
seems to be slightly overabundant in NGC\,6309, N may be deficient, and the
N/O abundance ratio is very low. Albeit multi-axis, 
NGC\,6309 is not a type\,I PN (V\'azquez et al. 2008). Chemical abundances in PNe are primarily 
related to the initial mass of the progenitor. Therefore, K\,4-37 and
NGC\,6309 should have evolved from progenitors of very different mass. The chemical abundances in NGC\,6309 
(Table\,3) and the models by Karakas (2010) suggest an initial mass of $\sim$1--1.2\,M$_{\odot}$ for its 
progenitor, while 4--6\,M$_{\odot}$ were estimated in the case of K\,4-37 (see
above). V\'azquez et al. (2008) noticed that the existence of the multi-axis 
structure of NGC\,6309 strongly contrasts with the idea that these kind 
of PNe should be associated with intermediate-mass progenitors (e.g., Corradi
\& Schwarz 1995). The striking structural similarities between K\,4-37 and NGC\,6309
strengthen this conclusion: stars with very different initial mass 
have been able to shape very similar PNe at the end of their lives. The
complexity of K\,4-37 and NGC\,6309 is compatible with that expected
from binary central stars and, in particular, from those having evolved
through a common envelope phase. It is known that post-common envelope binary
central stars tend to be associated with complex PNe that often show 
multiple structures and/or signs of collimated/focused outflows 
(Miszalski et al. 2009; Aller et al. 2015). In 
NGC\,6309, evidence (but not conclusive yet) exists for a possible F3V
companion, as suggested by the near-IR excess observed towards its central
star (Douchin et al. 2015). In K\,4-37 such an evidence does not exist. 
Nevertheless, given the complexity of the nebula, it is reasonable to propose 
a binary nature for the central star of K\,4-37. 

\section{Conclusions}

We have presented and analyzed narrow- and broad-band images, intermediate-
and high-resolution spectra, and WISE archive images of K\,4-37, a PN not analyzed 
before. The main conclusions of this work can be summarized as follows.

\begin{itemize}

\item K\,4-37 appears as a bipolar PN consisting of a bright equatorial torus,
  two main bipolar lobes, point-symmetric structures and off-axis
  deformations of the bipolar lobes. Its internal kinematics cannot be
  reproduced assuming an homologous expansion velocity law. The morphokinematical
  analysis allows us to disclose the existence of three distinct axes and
  additional particular directions in K\,3-37 that may be classified as
  a multi-axis PN. 

\item Very strong [N\,{\sc ii}], [S\,{\sc ii}] and relatively strong He\,{\sc
    ii} emission lines  are observed in the nebular spectrum. High He and N
  abundances, and a very high N/O abundance ratio ($\sim$2.32) are obtained. A
  progenitor of $\sim$4--6\,M$_{\odot}$ is estimated. The nebula presents a
  very low electron density indicating a highly evolved PN. 

\item A distance of $\sim$14\,kpc is estimated for K\,4-37, which is
  compatible with its highly evolved PN nature and a kinematical age of
  $\sim$10$^4$\,yr.  

\item The WISE image at 22\,$\mu$m reveals K\,4-37 to be surrounded by a very
  large elliptical detached shell with a size of $\sim$13$\times$8\,pc$^2$ and a
  kinematical age of $\sim$4.3$\times$10$^5$\,yr. This shell is 
  probably related to the mass ejected during the AGB phase of the progenitor 
of K\,4-37.

\item The elliptical morphology of the detached shell is not
  compatible with the spherical mass ejection expected in the AGB
  phase. Deformation of an originally spherical shell by
  the ISM magnetic field and/or the influence of a
  companion could explain the peculiar morphology. The formation of the nebula
  itself is better understood as caused by several focussed outflows at
  different directions and with different collimation degrees. 

\item Remarkable morphological similarites exist between K\,4-37 and
  NGC\,6309. However, chemical abundances largely differ in both PNe,
  indicating progenitors of very different initial mass. These results suggest
  that the initial mass of the progenitor has not played a crucial role in
  shaping these two PNe but their formation can be better attributed to a
  similar mass loss history of their central stars that, given the complexity
  of both PNe, are probably binary.  

\end{itemize}

\section*{Acknowledgements}
We are very grateful to our referee for his/her comments that have help to 
improve our analysis, interpretation and presentation of the results. We thank Calar Alto Observatory 
for allocation of director's discretionary time to this programme. We are very grateful
to the staff on Calar Alto for carrying out these observations. We thank the
staff of OAN-SPM for assistance during observations. We thank Wolfgang
Steffen for interesting discussions and help with {\sc shape}. This publication makes use of 
data products from the Wide-field Infrared Survey Explorer, which is a joint 
project of the University of California, Los Angeles, and the Jet Propulsion 
Laboratory/California Institute of Technology, funded by the National Aeronautics and 
Space Administration. LFM acknowledges partial support from Spanish MINECO grant
AYA2014-57369-C3-3-P (co-funded by FEDER funds). This project is supported by
UNAM-DGAPA-PAPIIT grant IN107914. Part of this paper was done during a stay of
LFM at the IA-UNAM (Ensenada,
M\'exico). He is very grateful to the people of the IA-UNAM for their warm 
hospitality and pleasant stay.







\bsp	
\label{lastpage}
\end{document}